\documentclass[aps,prl,twocolumn]{revtex4}
\usepackage{epsf}
\usepackage{times}
\begin{document}

\newcommand{\lessim}{\mbox{\tiny$\mbox{\normalsize$<$}\atop
\mbox{\normalsize$\sim$}$}}
\newcommand{\grtsim}{\mbox{\tiny$\mbox{\normalsize$>$}\atop
\mbox{\normalsize$\sim$}$}}

\title{Unconventional superconducting states induced in a ferromagnet  
by a $d$-wave superconductor}
\author{Zahra Faraii$^{1}$ and Malek Zareyan$^{1,2}$}

\affiliation{
$^{1}$Institute for Advanced Studies in Basic Sciences, 45195-159,
  Zanjan, Iran\\
$^{2}$ Max-Planck-Institute f\"ur Physik komplexer Systeme, 
N\"othnitzer Str. 38, 01187 Dresden, Germany}

\date{\today}

\begin{abstract}
We develop a quasi-classical theory for the 
superconducting proximity effect in a ballistic 
ferromagnetic layer in contact with a $d$-wave 
superconductor. In agreement with recent 
experiments we find that the density of states 
oscillates around the normal state value 
with varying the thickness of the ferromagnetic 
layer. We show that the phase, the amplitude, 
and the period of these oscillations 
depend on the orientation of the superconductor.
This effect reveals spatial oscillations and 
anisotropy of the induced superconducting 
correlations in the ferromagnet. 
\end{abstract}

\maketitle

Proximity effects in hybrid structures of 
superconductors and ferromagnets provide the
possibility for the controlled studies of the 
coexistence of ferromagnetism and 
superconductivity. Superconducting correlations 
penetrate into normal metals by means of a 
special scattering process at the 
normal metal-superconductor (NS) interface 
known as Andreev reflection \cite{Andreev}: 
an electron propagating in the normal metal 
can cross the NS boundary by reflecting as 
a hole and transferring a Copper pair into 
the superconductor. The Andreev reflected 
hole is correlated with the incident electron 
and carries information about the phase 
of the order parameter of the superconductor. 
The existence of correlated electron-hole 
pairs corresponds to a non-vanishing superconducting 
pair amplitude (order parameter), which gives to 
the normal metal the typical characteristics 
of the superconducting states \cite{Tink}. 
\par
The peculiarity of this proximity effect in 
ferromagnetic metals comes from the fact that
the Andreev reflection from a singlet pairing 
superconductor is accompanied by an inversion 
of the spin direction, and consequently the 
correlated electron-hole occupy opposite 
spin bands. The spin splitting exchange field 
of the ferromagnet causes a phase shift of 
the correlated Andreev electron-hole pairs, 
which results in a spatially oscillatory 
behavior of the induced pair amplitude \cite{paos}. 
Such an inhomogeneous superconducting 
state coexisting with ferromagnetism 
is strikingly similar to an FFLO state 
in a bulk superconductor with spin splitting
originally predicted by Fulde and Ferrell and by 
Larkin and Ovchinnikov \cite{FFLO}.
\par
One of the most interesting manifestations 
of the oscillations in the proximity pair 
amplitude is the oscillatory behavior of 
the density of states (DOS) in 
ferromagnet-superconductor bilayers
as function of the ferromagnetic layer 
thickness.  Kontos {\it et al} \cite{aprili01}
have observed this effect experimentally in 
thin ferromagnetic layers connected to a 
conventional $s$-wave superconductor. 
Theoretically, the superconducting proximity 
effect in ferromagnetic layers has been 
studied extensively \cite{paos,zbn02}. 
It has been shown \cite{zbn01} that the 
theory based on the quasi-classical 
formalism \cite{Eil} is in a quantitative 
agreement with the experiments. 
\par
In parallel, there also has been much 
attention to the proximity of a ferromagnet 
to  high-$T_c$ superconductors, which are 
widely believed to have a dominant $d$-wave 
pairing symmetry \cite{tsuei00}. Most of the 
studies have concentrated on the influence 
of the exchange field on the zero bias 
conductance peak (ZBCP) as the most remarkable 
feature in the tunneling spectroscopy of a 
junction between a high-$T_c$ superconductor 
and a normal metal \cite{hu94,tanaka95}. 
ZBCP is the result of a zero-energy Andreev
peak (ZEAP) in the DOS, which originates from 
change in the sign of the $d$-wave order parameter
under a $\pi/2$ rotation.
Many effects including a splitting of the 
ZBCP and the developing of a zero bias conductance 
dip (ZBCD) have been found and attributed 
to the exchange splitting \cite{larkin98,zhu00,stef02}.
\par
Most recently, Freamat and Ng \cite{Ng03} have 
performed tunneling spectroscopy measurement on 
multilayered junctions of $d$-wave superconductor 
and ferromagnets. They observed an oscillatory 
behavior in the tunneling conductance spectra 
with the thickness of the ferromagnetic 
domain. This effect has been considered
as the $d$-wave analog of the effect 
observed in \cite{aprili01} and explained 
as the signature for an FFLO state induced 
in the ferromagnetic layers  by the $d$-wave 
superconducting layer. However one would 
expect that the anisotropy and the sign 
change of the $d$-wave superconducting 
order parameter leads to an anisotropy 
and phase effects in the oscillation of 
the DOS, which distinguish it from the case 
of a conventional $s$-wave superconductor.
The goal of the present work is to study this 
effect theoretically. 
\par 
In this letter we study the proximity effect 
at the interface between a ferromagnetic metal 
and a $d$-wave superconductor. We model the 
ferromagnet as a thin ballistic film with 
rough boundaries which is connected to the 
superconductor through a disordered interface 
of finite transparency. We make use of the 
quasi-classical formalism \cite{Eil} to 
calculate the density of states in the 
ferromagnetic layer. 
In correspondence with the recent experiments, 
 we find that at sufficiently strong exchange 
field the DOS oscillates as a function of the 
thickness of the ferromagnetic layer.
We show that the phase, the amplitude and 
the period of the DOS oscillations are affected
by changing the orientation of the $d$-wave order 
parameter. This effect, which reveals the highly 
anisotropic nature of the induced 
superconducting correlations in the ferromagnet, 
can also be used as a further test of the $d$-wave 
scenario in high-$T_c$ superconductors. 
\begin{figure}
   \centerline{\hbox{\epsfxsize=2.in
                      \epsffile{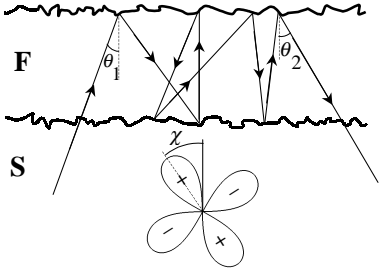}
                      }}
\caption{Sketch of a ferromagnetic layer (F) 
in connection with a $d$-wave superconductor 
 (S) . A typical classical trajectory is also 
shown. It starts at S and extends into F making
an angle $\theta_1$ with the interface normal 
and then, after several reflections from the 
insulator and FS-interface, returns into  S at 
an angle $\theta_2$.    
\label{fig1}}
\end{figure}
\par
We consider the proximity system shown 
in Fig. \ref{fig1}. A thin ferromagnetic 
layer (F) of thickness $d$ is contacted 
by a $d$-wave superconductor (S) on one 
side and covered on the other side by
an insulator. We consider a clean structure 
at both of the S and F sides of the 
FS-interface. At the same time the 
FS-interface itself may contain disorder 
and band mismatch which enhance 
backscattering of quasi-particles 
from this boundary. F is characterized 
by a mean exchange field $h$ which is 
assumed to be constant in the F-layer and 
vanishes in S. S is characterized by a 
$d_{x^2-y^2}$-wave order parameter of 
the form 
$\Delta _{d}(\theta)=\Delta\cos{[2(\theta-\chi)]}$, 
 where $\theta$ is the angle that the 
Fermi velocity ${\bf v}_{{\text F}}$ makes 
with the normal to the interface and 
$\chi$ is the angle between the 
crystallographic $a$ axis of S and 
the interface normal.  
For calculation we adopt the Eilenberger 
equations \cite{Eil} in the clean limit 
for this system. In the absence of spin-flip 
scattering in F, transport of quasi-particles
with spin $\sigma$ ($=\pm 1$) 
is described by an independent equation 
for the corresponding matrix Green's function 
$\hat{g}_{\sigma}(E,{{\bf{v}}_{{\text F}}},{\bf r})$, 
 which reads   
\begin{eqnarray}
\nonumber
&&-i{\bf{v}}_{{\text F}}{\bf \nabla}
  \hat{g}_{\sigma}(E,{{\bf{v}}_{{\text F}}},{\bf r})=
  [(E+\sigma h({\bf r}))\hat{\tau}_3\\
&&-i\hat\tau_2\Delta({\bf r},{\bf{v}}_{\text F}),
  \hat{g}_{\sigma}(E,{{\bf{v}}_{\text {\text F}}},{\bf r})].
  \label{Eil}
\end{eqnarray}
Here
$\hat{\tau}_i$ denote the Pauli matrices and
$\Delta (\bf{r},{{\bf{v}}_{{\text F}}})$ is the
superconducting order parameter,  
 which is taken to be real. We neglect the spatial 
variation of the order parameter close to the 
FS-interface and take 
$\Delta({\bf r},{\bf{v}}_{{\text F}})=
\Delta_{d}({\bf v}_{{\text F}})$
in all points of S. 
\par
Equation (\ref{Eil}) can be conveniently 
solved along a classical trajectory. A 
typical trajectory is shown in 
Fig. \ref{fig1}. It starts at  
S and extends into the F layer making an 
angle $\theta_1$ with the normal to the 
FS-interface. After several reflections 
from the insulator and the FS interface,
 which makes a path of total length $l$ 
in F, the trajectory returns into S at an 
angle $\theta_2$. The superconducting 
order parameter that an electron feels 
in the beginning and at the end of 
the trajectory are respectively 
$\Delta_d (\theta_1)$  and $\Delta_d (\theta_2)$.
We have solved Eq. (\ref{Eil}) and 
found that the DOS on a given trajectory 
only depends on $\theta_1$, $\theta_2$, 
and $l$ and given by
\begin{widetext}
\begin{eqnarray}
N(E,\theta_1,\theta_2,l)= 
\frac{N_0}{4}\sum\limits_{\sigma=\pm1}
  {\text {Re}}[{\text {tr}}\hat{\tau}_3
\hat{g}_{\sigma}(E,{{\bf{v}}_{{\text F}}},{\bf r})]
= \frac{N_0}{2}
\sum\limits_{\sigma=\pm1}
  {\text {Re}}\{-i
\tan {[k_\sigma l/v_{F} 
+\alpha(\theta_1)+
\alpha(\theta_2)+\eta(\theta_1,\theta_2)/2]}\},
\label{dostrj}
\end{eqnarray}
\end{widetext}
where $N_0$ is DOS at the Fermi level in the 
normal state,
$\alpha(\theta)=\arcsin{[E/|\Delta_d(\theta)|]}/2$, 
$k_{\sigma}=(E+\sigma h)/v_{F}$ and 
$\eta (\theta_1,\theta_2)=\pi[1-{\text {sign}}
(\Delta_d(\theta_1)\Delta_d(\theta_2))]/2$ 
is the $\pi$ phase shift resulting from the 
possible change in the sign of the order 
parameter at the beginning and the end of 
the trajectory.
\par 
The total density of states $N(E)$ is obtained 
by averaging the DOS per trajectory 
Eq. (\ref{dostrj}) over all the possible 
values of $\theta_1$,$\theta_2$, and $l$: 
\begin{equation}
 N(E)=\int dl d\theta_1 d\theta_2
 \bar{ p}(l,\theta_1,\theta_2)
 N(E,l,\theta_1,\theta_2),
\label{dosf}
\end{equation}
where ${\bar p}(l,\theta_1,\theta_2)$ is 
the distribution function of the trajectories. 
\par
To determine  ${\bar p}(l,\theta_1,\theta_2)$, 
 we model F as a weakly disordered thin layer 
bounded by  a rough surface and a rough 
FS-interface with a constant transparency $t$. 
We introduce $g(\theta_1,\theta_2)$ as the 
correlation function between the incoming and 
outgoing directions $\theta_1$ and $\theta_2$. 
 This function determines the specularity of 
the scattering from the boundaries. For 
perfectly flat boundaries with purely specular 
scattering, $\theta_1$ and $\theta_2$ are 
completely correlated and 
$g\sim \delta(\theta_1+\theta_2)$.
 For the boundaries with strong roughness the 
dominantly diffusive scattering destroys any 
correlation between $\theta_1$ and $\theta_2$, 
 and $g(\theta_1,\theta_2)$ is constant. 
In the general case of an arbitrary strength 
of the roughness  we assume 
$g=C^{-1} {\text {exp}}(-(\theta_1+\theta_2)^2
/(\pi z)^2)$, where $C=\int_{-\pi/2}^{\pi/2}
d\theta_1d\theta_2g(\theta_1,\theta_2)$ is a 
normalization factor and $z$ measures the 
degree of the roughness. The limits of 
$z\ll 1$ and $z\sim 1$ correspond to weak 
and strong roughness respectively.
 As we will see below the parameter $z$ play 
a crucial role to reveal the anisotropic nature 
of the induced superconducting correlations 
in the F-layer.    
\par
The full distribution function 
${\bar p}(l,\theta_1,\theta_2)$ can be  written 
in terms of $g(\theta_1,\theta_2)$  as  
\begin{eqnarray}
\nonumber
 && \bar{p}(l,\theta_1,\theta_2)=g(\theta_1,\theta_2)[
 t\delta (l-\frac{d}{\cos{\theta_1}}
 -\frac{d}{\cos{\theta_2}})\\
 && +rp(l-\frac{d}{\cos{\theta_1}}-\frac{d}{\cos{\theta_2}})],
\label{pltt}
\end{eqnarray}
where $p(l)$ is the length distribution of the 
trajectories regardless of the values of 
$\theta_1$ and $\theta_2$, and $r=1-t$. 
For $t=1$ the delta function in Eq. (\ref{pltt})  
assure that the total length of the trajectory 
is given by $l=d/\cos{\theta_1}+d/\cos{\theta_2}$.  
In the limit of small transparency of the 
FS-interface $t\ll 1$, $p(l)$ can be approximated 
by the exponentially decaying function \cite{zbn02}  
\begin{equation}
  p(l)=\frac{1}{{\bar l}}e^{-l/{\bar l}}\Theta (l-2d), 
  \label{pl}
\end{equation}
where ${\bar l}\simeq 2d\ln(\ell_{{\text imp}}/d)/t
=2d_t$ is the mean trajectory length in the limit 
of small $d/\ell_{{\text imp}}$ ($\ell_{{\text imp}}$ 
is the elastic mean free path). 
\par
Combining  Eqs.  (\ref{dostrj}), (\ref{dosf}), 
(\ref{pltt}), and (\ref{pl}) we find the 
following result for the total DOS  
\begin{equation}
  N(E)=\frac{N_0}{2}\sum_{\sigma=\pm 1}
  \sum_{n=-\infty}^{\infty}[t+rp(2nk_{\sigma}d)]J_n,
  \label{dost}
   \end{equation}
where $p(k)=e^{-i2kd}/(ik{\bar l}+1)$ 
is the Fourier transform of $p(l)$ and  
\begin{eqnarray}
\nonumber
J_n(k)=\int_{-\pi/2}^{\pi/2}d\theta_1d\theta_2
g(\theta_1,\theta_2)
f_n(\theta_1)f_n(\theta_2)
e^{-in\eta(\theta_1,\theta_2)},
 \label{zn}
\end{eqnarray}
\begin{eqnarray}
\nonumber
f_n(\theta)=
\left \{ 
\begin{array}{l@{{\text {for}}}r}
e^{-i2nk_{\sigma}d/\cos{\theta}+i2n\phi(\theta)}\quad &
\quad |E|\leq|\Delta_d(\theta)| \\
e^{-i2nk_{\sigma}d/\cos{\theta}-2|n|\beta(\theta)} \quad&
\quad |E| > |\Delta_d(\theta)|
\end{array} \right. .
 \label{ff}
\end{eqnarray}
Here $\phi(\theta)=\arccos{[E/|\Delta_d(\theta)|]}/2$ 
is the Andreev phase and 
$\beta(\theta)=
{\text {acosh}}{[E/|\Delta_d(\theta)|]}/2$.
\par
\begin{figure}
   \centerline{\hbox{\epsfxsize=3.37in
                      \epsffile{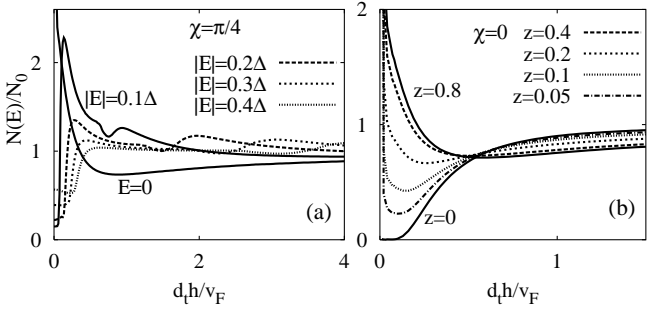}
                      }}
\caption{
Effect of weak exchange fields on the 
density of states when the transparency 
of the FS-interface is small (here
$t=0.1$). Normalized DOS $N(E)/N_0$ 
as a function of $d_th/v_{{\text F}}$ 
($d_t\sim d/t$) at (a) different energies 
when $\chi=\pi/4$ and $z=0.5$, and 
(b) zero energy when $\chi=0$ and for 
different strength of the roughness of 
boundaries $z$. The exchange field 
suppresses the typical 
superconducting features of the DOS 
(the zero energy Andreev peak when 
$\chi=\pi/4$,  and also when $\chi=0$ 
if the boundaries are rough, and the 
vanishing DOS at $E=0$ when the 
boundaries are specular and $\chi=0$).
\label{fig2}}
\end{figure}
Equation (\ref{dost}) expresses the total 
DOS in terms of known functions. In real 
samples  the F-layer is expected to have a 
nonuniform thickness due to the large scale 
roughness of the boundaries. To take this 
into account we average (\ref{dost}) 
over a Gaussian distribution of the 
thicknesses around a mean value $d$. This 
leads to a smearing of the sharp features 
in the DOS which arise from the lower cutoff 
in the length distribution. The qualitative 
behavior,  however, will not be changed. 
In practice, we have taken the width of the 
distribution to be of order 10\%, which 
corresponds to the conditions of the 
experiments \cite{aprili01,Ng03}.      
\par
Let us start by analyzing the effect of a 
weak exchange field on the proximity DOS of 
a thin F-layer with $d\ll v_{{\text F}}/\Delta$ and 
$t\ll 1$. For the orientation $\chi=\pi/4$ the 
main feature in the DOS of a normal layer 
($h=0$) is a sharp ZEAP for any strength of the 
roughness $z$. In Fig. \ref{fig2}a the 
normalized DOS $N(E)/N_0$ versus 
$d_th/v_{\text {F}}$ is plotted for different 
energies and values $\chi=\pi/4$, $z=0.5$. 
The effect of the exchange field is to split 
the ZEAP into distinct peaks at finite energies. 
Increasing $h$ further leads to the shifting 
of the splitted peaks toward higher energies, 
 which is associated with decreasing the height 
of these peaks. The DOS at $E=0$ decreases with 
$h$ and goes through a minimum as 
$d_th/v_{\text {F}}$ is increasing. Thus the 
exchange field can suppress the ZEAP 
in the DOS and develop a dip at $E=0$.  
This will lead to a ZBCD in the conductance 
spectra at low temperatures, which also has 
been observed in the experiments \cite{larkin98}. 
\par
For the orientation $\chi=0$ the ZEAP is absent 
for perfectly specular boundaries ($z=0$). The ZEAP,  
 however, appears when $z$ is finite. This is the 
result of diffusive scattering caused by the roughness 
at the boundaries, which allows for the sign change 
of the superconducting order parameter at two sides 
of a sizable fraction of trajectories even when 
$\chi=0$ \cite{fogel97}. 
In Fig. \ref{fig2}b  we plot the normalized 
DOS $N(0)/N_0$ at $E=0$  versus 
$d_th/v_{\text {F}}$ when $\chi=0$ and for 
different values of $z$. For the normal layer 
 ($h=0$) the height of the ZEAP increases with 
increasing $z$. The effect of the exchange 
field for a given value of $z$ is to suppress 
the superconducting features (zero DOS for 
$z=0$ and ZEAP for finite $z$ ) in the DOS.
At higher exchange fields when 
$d_th /v_{\text {F}} \grtsim 1$, the zero 
energy DOS for all values of $z$ approach 
each other and becomes close to the normal 
state values $N_0$. 
\par    
\begin{figure}
   \centerline{\hbox{\epsfxsize=3.2in
                      \epsffile{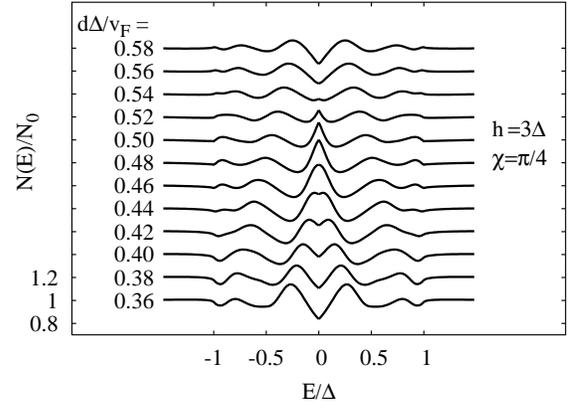}
                      }}
\caption{
Oscillations of the proximity density of 
states with the F-layer thickness $d$.
Dependence of the normalized DOS $N(E)/N_0$ 
on $d$ when the interface is highly transparent 
($t=1$) and $h=3\Delta$, $\chi=\pi/4$, $z=1$.
\label{fig3}}
\end{figure}
At strong exchange fields when 
$dh/v_{\text{F}}\gtrsim 1$, the DOS at all 
energies oscillates around the normal state 
values $N_0$ as a function of $dh/v_{\text{F}}$. 
This oscillatory behavior of induced 
superconducting correlations has been observed 
in the experiments \cite{Ng03}. Fig. \ref{fig3} 
represents these oscillations resulting from our
calculations. The energy dependence of the 
normalized DOS $N(E)/N_0$ is plotted for different 
values of the F-layer thickness when the 
FS-interface is highly transparent ($t=1$), 
$h=3\Delta$ and $\chi=\pi/4$. We have taken 
$z=1$, modeling rough boundaries. In this 
case the period of the DOS oscillations 
is roughly $\pi/4$.
\par
\begin{figure}
\centerline{\hbox{\epsfxsize=3.in
                      \epsffile{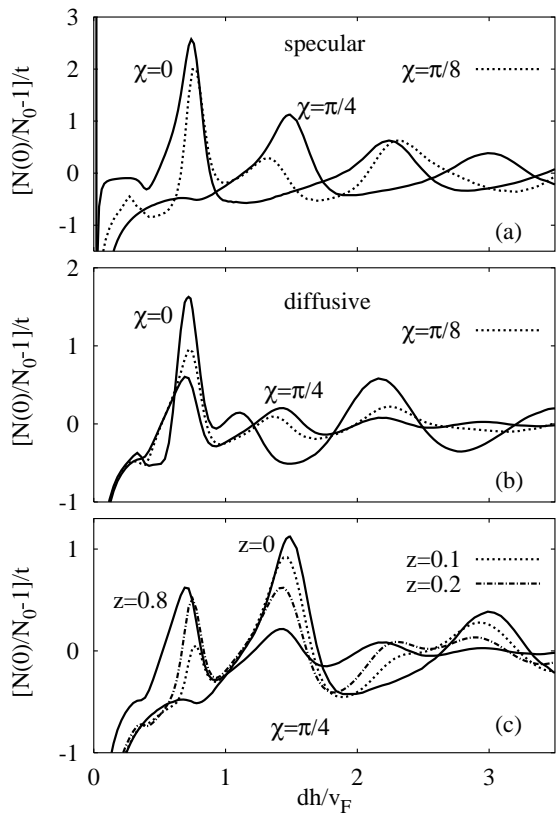}
                      }}

\caption{ Effect of the $d$-wave symmetry 
on the oscillations of the DOS. 
The reduced DOS at zero energy $[N(0)/N_0-1]/t$ 
versus $dh/v_{\text {F}}$ for 
different orientations $\chi$ when the boundaries 
are (a) specular ($z=0$)  and (b) diffusive ($z=1$).
In both cases the amplitude and the phase of 
the oscillations are modulated by varying $\chi$. 
For diffusive boundaries the period of the 
oscillations is also different for two cases 
of $\chi=\pi/4$ and $\chi=0$. (c) shows the 
transition from the specular to the diffusive 
case by varying $z$ for fixed $\chi=\pi/4$ .   
\label{fig4}}
\end{figure}
 
Let us now analyze the effect of the $d$-wave 
symmetry of the superconducting order 
parameter on the oscillations of the DOS 
in F. For this we examine the oscillatory 
behavior of the DOS at the Fermi level 
($E=0$) for different orientations of S.
We consider the case of $t\ll 1$ where for 
$h\gtrsim v_{\text{F}}/d $ the deviations 
of the DOS from the normal state value $N_0$ 
are of order $t$.  
In Fig. \ref{fig4}a  we have plotted the 
reduced DOS $[N(0)/N_0-1]/t$ versus 
$dh/v_{\text{F}}$ for the case of specular 
boundaries ($z=0$) at different $\chi$. 
The variation of  $\chi$ affects the phase 
and the  amplitude of the oscillations. 
Varying $\chi$ from $0$ to $\pi/4$, the 
phase is shifted by $\pi$ and the 
amplitude is decreased. The $\pi$ phase 
shift originates from the sign change 
of the order parameter, which occurs 
for all the trajectories in specular 
boundaries if $\chi=\pi/4$.
We note that the period of the oscillations 
is almost the same for different $\chi$ 
and equals to $\pi/2$, the period for the 
case where S has a $s$-wave order parameter 
\cite{zbn01}. 
\par
The effect of an anisotropic order parameter 
is more pronounced for the case of diffusive 
boundaries. This is shown in Fig. \ref{fig4}b
where the DOS oscillations at $E=0$ are presented 
for the case of  $z=1$. Now, in addition to 
the variation of the phase and the amplitude,
 by changing the S orientation from $\chi=0$ to  
$\chi=\pi/4$ the period of the oscillations 
is also changed. Fig. \ref{fig4}c shows how 
the change in the period occurs  when the 
strength of the roughness $z$ varies for 
$\chi=\pi/4$.
We can understand these effects by noting 
that the spatially oscillating order parameter
in the F-layer has a direction-dependent 
amplitude and phase resulting from the 
anisotropy and the sign change of the $d$-wave 
order parameter of S. The change in 
the phase, amplitude and the period of the 
DOS oscillations is the result of unconventional 
induced superconducting correlations    
\par
In conclusion, we have investigated theoretically  
the superconducting proximity effect in 
a thin ferromagnetic layer in contact with 
a $d$-wave superconductor. In correspondence 
with recent experiments \cite{Ng03}  we have 
found that at sufficiently strong exchange fields 
the density of states in the ferromagnet 
oscillates around the normal state value 
as a function of its thickness.  
The phase, the amplitude, and the period of 
the oscillations depend on the orientation 
of the superconductor. This direction-dependence 
is the signature of an unconventional oscillatory 
superconducting state induced in the ferromagnet 
by the proximity to the $d$-wave superconductor.
\par
We thank M. Freamat, P. Fulde, Yu. V. Nazarov, and H. Schomerus
for useful discussions.  

\end{document}